\documentclass[journal]{IEEEtran}

\usepackage{cite}
\usepackage[Symbol]{upgreek}

\usepackage{amsthm,amsmath,amssymb,amsfonts,mathrsfs,mathtools,dsfont,mathrsfs,bm,bbm}
\DeclareMathAlphabet{\mathpzc}{OT1}{pzc}{m}{it}
\DeclareSymbolFontAlphabet{\amsmathbb}{AMSb}%
\usepackage{csvsimple,booktabs}
\usepackage{float,enumitem}
\usepackage{graphicx,epsfig,xcolor}
\usepackage{balance}
\usepackage[hyphens]{url}
\usepackage[breaklinks, colorlinks, linkcolor=purple, anchorcolor=purple, citecolor=purple, urlcolor=purple]{hyperref}
\usepackage{multirow}
\usepackage[T1]{fontenc}

\usepackage{amsmath}
\usepackage[cmintegrals]{newtxmath}
\usepackage{mwe}
\usepackage{capt-of}
\usepackage{amsmath,amsfonts}
\usepackage{algorithmic}
\usepackage{multicol}  
\usepackage{threeparttable}  
\usepackage{booktabs}  
\usepackage{array}
\newtheorem{theorem}{\bf Theorem}
\newtheorem{proof}{\bf Proof}

\usepackage[ruled,vlined,linesnumbered]{algorithm2e}
\makeatletter
\newcommand{\removelatexerror}{\let\@latex@error\@gobble}
\makeatother
\usepackage{subfigure}
\usepackage{subcaption}
\usepackage{psfrag}
\usepackage{xcolor}

\allowdisplaybreaks[4]
 \usepackage{enumerate}

\ifCLASSINFOpdf
\else
\fi
\makeatletter
\patchcmd{\@maketitle}
  {\addvspace{0.5\baselineskip}\egroup}
  {\addvspace{-1.6\baselineskip}\egroup}
  {}
  {}
\makeatother
\begin{document}
%
\renewcommand{\baselinestretch}{0.885}
\title{Optimal Operation of Distribution System Operator and the Impact of Peer-to-Peer Transactions}
%
%
%

\author{Hanyang~Lin, Ye~Guo,~\IEEEmembership{Senior Member,~IEEE,}
        Firdous~Ul~Nazir,~\IEEEmembership{Senior Member,~IEEE,} Jianguo~Zhou,~\IEEEmembership{Member,~IEEE,}  Chi~Yung~Chung,~\IEEEmembership{Fellow,~IEEE}
        and Nikos~Hatziargyriou,~\IEEEmembership{Life Fellow,~IEEE}
\thanks{This work was supported in part by the National Natural Science Foundation of China under grant 52377105. (Corresponding author: Y. Guo: guo-ye@sz.tsinghua.edu.cn)}
\thanks{H. Lin, Y. Guo and J. Zhou are with Tsinghua-Berkeley Shenzhen Institute, Shenzhen International Graduate School, Tsinghua University, Shenzhen, 518071, China. F. Ul Nazir is with the Department of Electrical and Electronic Engineering, Glasgow Caledonian University, Glasgow, G4 0BA, U.K. C. Y. Chung is with the Department of Electrical and Electronic Engineering, Hong Kong Polytechnic University, Kowloon, 999077, Hong Kong. N. Hatziargyriou is with the School of Electrical and Computer Engineering, National Technical University of Athens, 157 73 Athens, Greece.}}
\maketitle

\begin{abstract}
Peer-to-peer (P2P) energy trading, commonly recognized as a decentralized approach, has emerged as a popular way to better utilize distributed energy resources (DERs). In order to better manage this user-side decentralized approach from a system operator's point of view, this paper proposes an optimal operation approach for distribution system operators (DSO), comprising internal prosumers who engage in P2P transactions. The DSO is assumed to be a financial neutral entity, holding the responsibility of aggregating the surplus energy and deficit demand of prosumers after their P2P transactions while dispatching DERs and considering network integrity. Impacts of P2P transactions on DSO's optimal operation have been studied. Results indicate that energy matching P2P trading where only the total amount of energy over a given period of time is defined may affect quantities of energy exchanged between the DSO and the wholesale market, but not internal dispatch decisions of the DSO. Different levels of real-time power consistency may lead to different total surpluses in the distribution network. For the real-time power matching P2P trading, as a special case of energy matching P2P trading, the provided energy and total surplus are not affected. In other words, DSO can safely ignore P2P transactions if they follow the format defined in this paper. Case studies verify these conclusions and further demonstrate that P2P trading will not affect physical power flow of the whole system, but the financial distribution between the DSO and prosumers.
\end{abstract}

\section*{Nomenclature}
\addcontentsline{toc}{section}{Nomenclature}
\begin{description}[leftmargin=6.5em,style=nextline]
  \item[Notations]
  \item[$T$] Time horizon.
  \item[$E$] Energy market. 
  \item[$R$] Reserve market.
  \item[$g$] Thermal generators.
  \item[$S$] Storage systems.
  \item[$D$] Demand side management.
  \item[$PV$] Photovoltaic units.
  \item[$i$] Index of prosumer.
  \item[$s$] Index of scenario.
  \item[$t$] Index of time period.
  \item[$ex$] Index of exchanging power.
  \item[$k,j,l$] Index of bus.
  \item[$kj$] Index of branch.  
  \item[$Nb$] Set of all buses.
  \item[$Nl$] Set of all branches.
  \item[$N_{p}$] Set of all prosumers.
  \item[$N_{s}$] Set of all scenarios.
  \item[Parameters]
  \item[$\omega_{s}$] Probability of scenario $s$.
  \item[$\pi^{E}_{t}/\pi^{R}_{t}$] Energy/reserve market price at time $t$. 
  \item[$\alpha^{g}_{i}/\beta^{g}_{i}$] Generation cost coefficients of thermal generator with prosumer $i$.
  \item[$\alpha^{S}_{i}$] Degradation cost coefficients of storage system with prosumer $i$.
  \item[$\alpha^{U}_{i}/\beta^{U}_{i}$] Cost coefficients of the utility of prosumer $i$.
  \item[$P^{PV}_{s,i,t}$] Predicted output of the PV unit with prosumer $i$ at time $t$ in scenario $s$. 
  \item[$\underline{P}^{g}_{i},\overline{P}^{g}_{i}$] Minimum/maximum generation output of thermal generator with prosumer $i$.  
  \item[$P^{d}_{s,i,t}$] Predicted load of prosumer $i$ at time $t$ in scenario $s$. 
    \item[$P^{B}_{i,t}$] Generation output for P2P trading of prosumer $i$ at time $t$.
  \item[$MSR_{i}$] Ramping capability for reserve of thermal generator with prosumer $i$.   
  \item[$RU_{i}/RD_{i}$] Ramping up/down limit of thermal generator with prosumer $i$.   
  \item[$\overline{P}^{D}_{i,t}$] Adjustment capacity of flexible demand of prosumer $i$ at time $t$.  
  \item[$\overline{P}^{S}_{i,t}$] Maximum charge and discharge rate of storage system with prosumer $i$.
    \item[$\eta_{i}$] Charge and discharge efficiency of storage system with prosumer $i$.
  \item[$\overline{Q}_{i}$] Installed capacity of storage system with prosumer $i$.  
  \item[$Q^{0}_{i}$] Minimum capacity requirement of storage system with prosumer $i$.  
  \item[$\overline{E}_{ex}$] Maximum exchanging power between DSO and TSO at substation node.
  \item[$r_{kj},x_{kj}$] Series resistance/reactance of the branch between buses $k$ and $j$.
  \item[$z_{kj}$] Series impedance of the branch between buses $k$ and $j$.
  \item[$Gs_{kj},Bs_{kj}$] Shunt conductance and susceptance of the branch between buses $k$ and $j$.
  \item[$y^{m}_{kj}$] Shunt admittance of the branch between buses $k$ and $j$.
  \item[$\underline{V}_{j}/\overline{V}_{j}$] Minimum/maximum nodal voltage magnitude limit of node $j$. 
  \item[$\overline{S}_{kj}$] Maximum branch power magnitude limit of the branch between buses $k$ and $j$.   
  \item[Variables]
  \item[$E_{t}/R_{t}$] Energy/reserve provided by DSO to the wholesale market at time $t$.
  \item[$P^{g}_{s,i,t}$] Generation output of thermal generator with prosumer $i$ at time $t$ in scenario $s$. 
  \item[$P^{c}_{s,i,t}/P^{dis}_{s,i,t}$] Charging/discharging power of storage system with prosumer $i$ at time $t$ in scenario $s$.   
  \item[$P^{D}_{s,i,t}$] Load curtailment of flexible demand of prosumer $i$ at time $t$ in scenario $s$.     
  \item[$P^{E}_{s,i,t}$] Power output for energy market of prosumer $i$ at time $t$ in scenario $s$. 
  \item[$P^{R}_{s,i,t}$] Upward reserve capacity of thermal generator with prosumer $i$ at time $t$ in scenario $s$.   
  \item[$Q^{S}_{s,i,t}$] Energy state of storage system with prosumer $i$ at time $t$ in scenario $s$. 
  \item[$p_{s,j,t}/q_{j,t}$] Nodal active/reactive power extraction at node $j$ at time $t$ in scenario $s$. 
  \item[$p_{s,kj,t}/q_{kj,t}$] Active/reactive branch flow of the branch between buses $k$ and $j$ at time $t$ in scenario $s$.   
  \item[$S_{s,kj,t}$] Complex power flow of the branch between buses $k$ and $j$ at time $t$ in scenario $s$.   
  \item[$\ell_{s,kj,t}$] Squared current magnitude of the branch between buses $k$ and $j$ at time $t$ in scenario $s$.     
  \item[$v_{s,j,t}$] Squared voltage magnitude of bus $j$ at time $t$ in scenario $s$. 
 \end{description}
%
\IEEEpeerreviewmaketitle

\section{Introduction}

\subsection{Background}

\IEEEPARstart{D}{istributed} energy resources (DERs) including distributed generation, storage systems and smart appliances have gained widespread recognition as pivotal tools in achieving carbon neutrality. Moreover, the integration of DERs in power distribution systems has reshaped the role of the system operator and users. For example, traditional distribution systems are in the transition to active distribution networks (ADN), which strengthens the interaction between distribution system operators (DSO) and transmission system operators (TSO)\cite{chen2024multi}, and brings more possibilities to balance the system load\cite{LI2024122134}. This transition also enables a bi-directional energy and reserve interchange between the DSO and the wholesale electricity market, different from the traditional manner where the DSO only purchases the energy needed, aiming to realize the economic benefits of market participation and provide support for the transmission systems\cite{zhang2016real}. 

At the same time, the rapid growth of DERs' penetration makes traditional passive consumers become productive consumers (prosumers). It also facilitates the emergence of the peer-to-peer (P2P) energy trading platform, where prosumers can share surplus energy with their neighbouring peers through P2P transactions. No doubt that the potential benefit that prosumers can reap from P2P transactions would be substantial. For example, they have more choices to handle their surplus energy or deficit demand, and therefore can reduce their operational costs\cite{lei2022shareholding} and search for higher revenue. 

The DSO's optimal dispatching and market participation can be regarded as a centralized approach utilizing DERs, while the P2P energy trading is a decentralized way specifically designed for users' DERs. The value streams provided by both methods from different entities are important. However, P2P energy transactions would bring several concerns at system operators' view as prosumers do not consider the network integrity\cite{tushar2018peer}. Therefore, the combination of DSO's optimal dispatching and P2P energy trading can address these issues by using top-down strategies from DSOs\cite{morstyn2018using}.

\subsection{Related Works}

There have been several studies focusing on the utilization of DERs through DSO's optimal operation, P2P energy trading and their combinations. For the DSO's optimal operation, one important problem is to provide energy and reserve in the wholesale market through dispatching DERs. DSO in \cite{8091006} determines the amount of energy and reserve through a profit model at the day-ahead stage to maximize its profit. In \cite{kalantar2019characterizing}, the range of the active and reactive power reserves by ADNs is determined considering load forecast errors, generation output uncertainty and operating constraints of DERs. An integrated dispatching model of DERs was proposed in \cite{chen2018forming}, which optimally decides the DSO's provided energy and reserves in the wholesale market by aggregating DERs. However, the above works did not consider the security constraints of the network. An optimal decision-making model was proposed in \cite{zhou2019distributionally} for the DSO to participate in the joint energy-reserve market considering uncertainties of load and wind speed considering network securities. In \cite{mohan2015efficient}, DSO and microgrids are allowed to have energy and reserve transactions by utilizing the network and participating in the wholesale market together. 

The P2P energy trading may raise challenges for network integrity because prosumers are unaware of network operational constraints\cite{guerrero2018decentralized}. There is a body of works based on the construction of the P2P platform without violating the network’s security limits. One important direction aims at delivering a non-congested energy trading throughout the entire system while keeping the network loss at a minimum level\cite{baroche2019exogenous}. In \cite{kim2019p2p}, a P2P transaction scheme considering power losses was proposed, where relevant costs are counted in each transaction based on network utilization. Authors in \cite{nikolaidis2018graph} proposed a topology-based loss allocation method for P2P transactions in distribution networks. 

Therefore, there are studies focusing on the interaction of DSO's operation and P2P trading. Authors in \cite{9591485} propose a hierarchical coordination method for charging stations and DSO to make full use of the EVs' charging flexibility, considering P2P trading between EVs and stations. The DSO checks the system security with given pattern off P2P tradings. In \cite{9235517}, a two-level network-constrained P2P trading for microgrids was proposed, where microgrids can trade energy with each other, and DSO reconfigures the network considering P2P trading. Authors in \cite{botelho2022integrated} integrated the P2P energy and reserve trading into the distribution system through the operation decision of DSO. A DSO-prosumer cooperated scheduling framework was proposed in \cite{9810546}, where the power flow can be optimized by incorporating the P2P energy trading among prosumers into DSO's dispatching. However, these studies are not mainly focusing on the DSO's market participation in the wholesale market considering P2P trading.

In fact, the impact of P2P transactions on DSO's operation is still an open question. Intuitively, some believe that P2P tradings are only financial and do not affect DSO's physical operation, while many studies \cite{guerrero2018decentralized,9803219,baroche2019exogenous,9235517} believe that P2P tradings may affect the DSO's operation, leading to possible voltage violations and line congestions.

\subsection{Contributions}
In this paper, a stochastic optimal dispatch model for the DSO considering P2P tradings with estimated real-time uncertainties is developed. Both energy contract and power contract, representing two types of P2P transactions, are modelled within the proposed optimization. After internal P2P transactions between prosumers with their own DERs at different nodes, the DSO optimally dispatches the prosumers' DERs considering network integrity. Subsequently, it can enable local resource participation in wholesale markets and compensate resources for grid services \cite{tapio2023ferc} by providing energy and reserve in the wholesale market truthfully according to its operation model.

Based on the relaxed convex optimal dispatch model, it is theoretically proven that energy matching P2P transactions, as per the definition proposed in this paper, have no impact on the DSO's optimal dispatch but its trading decisions. We also reveal that P2P trading will not affect the physical power flow of the whole system. Instead it does bring more surplus for the prosumers who engaged in P2P trading and will decrease the total surplus of DSO when there exist real-time power mismatches. Furthermore, power matching contract will not affect DSO's trading decisions either. Moreover, numerical tests verify these conclusions based on the original non-convex model.

The rest of this paper is organized as follows. Section II presents the framework considered in this paper. Then, the model of DSO's optimal operation considering P2P trading is formulated in Section III. The impacts of P2P trading on the DSO's optimal operation are analytically presented in Section IV. Numerical results are provided in Section V, and conclusions are drawn in Section VI.
\begin{figure}[ht!]
  \centering
  \includegraphics[scale=0.55]{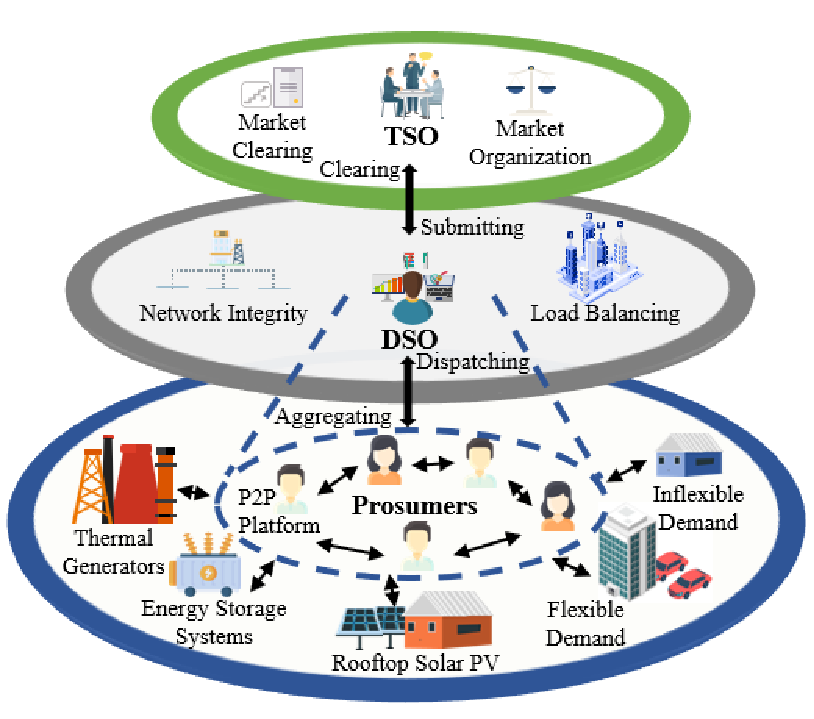}
    \captionsetup{singlelinecheck=off}
  \caption{Hierarchy and relationship among different subjects}
  \vspace{-0.5cm}
\end{figure}

\section{Framework}
This paper focuses on the DSO who bridges prosumers and the wholesale market operator, or the TSO. The DSO is responsible of managing all internal prosumers with certain inelastic demands, as shown in Fig. 1. Prosumers also posses four types of DERs: thermal generators, rooftop solar photovoltaic (PV) units, energy storage systems and curtailable loads. P2P trading may happen between prosumers located at different nodes in the distribution system, and without loss of generality, we assume that one node only has one prosumer. Given a certain pattern of P2P transactions, the DSO aggregates all the prosumers to submit quantities of hourly demand or supply for energy and reserve truthfully to the wholesale market operator. Moreover, DSO needs to consider several crucial responsibilities in its optimal operation model, including balancing the local demand and dispatching the prosumers' DERs considering distribution network operation limits. In this paper, without loss of generality, we consider that the wholesale market includes the energy market and upward spinning reserve market and omit the downward reserve in the reserve market.

\section{Problem Modelling}
\subsection{Modelling of P2P Transactions and Assumptions}
Two types of P2P trading are considered: energy contract and power contract. For example, consider a P2P transaction between a seller with a PV unit and a consumer, whose real-time PV output and load power are shown in Fig. 2. The energy contract P2P trading is defined based on the total energy over a certain period, which means that the seller’s daily energy generation is the same as the buyer’s daily consumption (total 20 MWh in this example). In other words, the energy contract only specifies that total amounts of power selling and buying are the same over a given time period, because the real-time power depends on trading prices, energy production and consumption which are unknown. While power contract P2P trading is defined as their common part, i.e. the purple curve from hours 8-19 in Fig. 2. In essence, a power contract restricts that amounts of power selling and buying should be equal to each other at any time.  

\begin{figure}[ht!]
  \centering
  \includegraphics[scale=0.45]{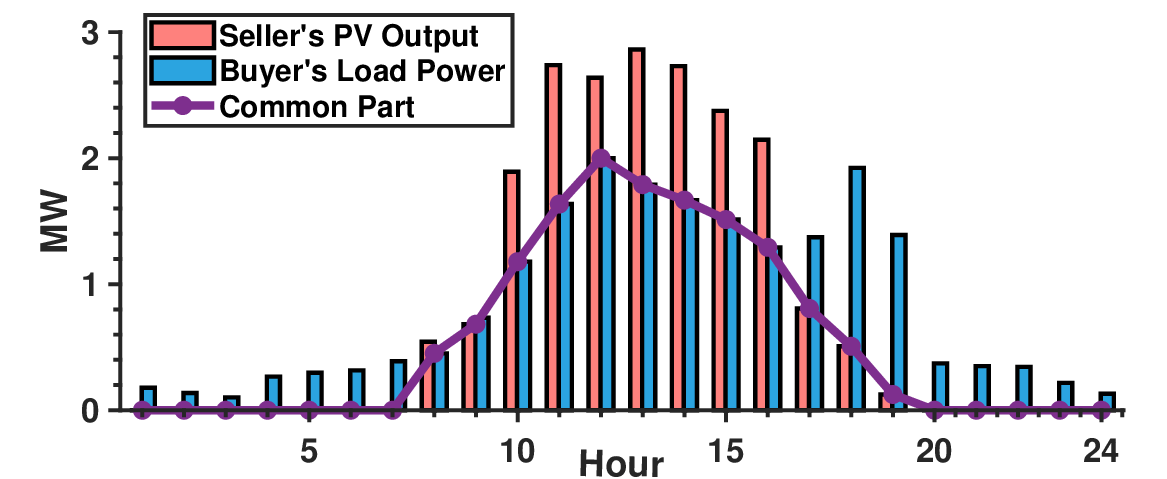}
    \captionsetup{singlelinecheck=off}
  \caption{PV output and load power}
  \vspace{-0.3cm}
\end{figure}

Each energy contract P2P trading $j$ is defined as $\{B_{j}, S_{j}, Q_{Bj}, Q_{Sj}, T_{j}\}$, which are the index of buying/selling prosumer, the vector of trading power of buyer/seller in P2P transaction $j$ in all time intervals, the vector of transaction time periods, respectively. Denote $T_{j,\tau}$ as the $\tau^{th}$ element of $T_{j}$. Then, for any $T_{j,\tau}$ in $T_{j}$ the trading power of buyer and seller of P2P transaction $j$ must satisfy
\begin{equation}
      \sum_{t\in T_{j,\tau}} Q_{Bj,t}=\sum_{t\in T_{j,\tau}} Q_{Sj,t},   
\end{equation}
where $Q_{Bj,t}/Q_{Sj,t}$ is the $t^{th}$ element of $Q_{Bj}/Q_{Sj}$. Then we can calculate the net P2P trading quantity of prosumer $i$ as
\begin{align}
    P^{B}_{i}= \sum_{\{j|S_{j}=i\}} Q_{Sj} - \sum_{\{j|B_{j}=i\}} Q_{Bj}.
\end{align}

The power contract P2P trading is a special case of energy contract P2P trading with the following condition:
\begin{align}
    Q_{Bj,t}=Q_{Sj,t}
    \vspace{-0.3cm}
\end{align}
$\forall t \in T$. The P2P trading mentioned below is energy contract P2P trading, if not specifically mentioned.

Since P2P transactions are considered as given parameters, in the optimal operation model of the DSO, it is sufficient to employ the vector of $P_i^B$ to present the contribution from all P2P transactions of prosumer $i$. However, it should be noticed that if all P2P transactions are power contracts, there is $\sum_{i} P_i^B =0$. 

In this section, a scenario-based optimal operation model for the DSO considering P2P trading between prosumers is formulated. We consider that prosumers have already decided their P2P trading contracts and reported the trading power to the DSO. Therefore, DSO considers P2P trading quantities as known parameters in constraints of its optimal operation model while omitting P2P trading prices which are not part of the DSO's decision. 

In addition, the following assumptions are made to develop our model for the optimal operation of the DSO:

i) Scenarios are generated with given probabilities using Monte Carlo method to model uncertainties in the prediction of PV generation and load power. 

ii) Given predicted market prices, the DSO submits the quantities of energy and reserve to the wholesale market truthfully according to its operation model. 

iii) Quantities of energy and reserve in the wholesale market are assumed to be independent to scenarios, which are decisions in the ex-ante stage. Its optimal dispatch of resources, however, are ex-post decisions that depend on the realization of uncertain variables.

iv) Each prosumer possesses all four types of resources as in section II, while a zero capacity means the prosumer does not have such a type of resources. Their PV units are set in each maximum power point tracking (MPPT) mode.

\subsection{Objective Function}
With a given set of P2P transactions, denoted by the P2P net trading quantity of prosumers, $P^{B}$, the DSO aggregates all prosumers and optimizes the total surplus defined by

\begin{align}
 \vspace{-0.3cm}
 &\underset{ \mathcal{O}}{\rm max}: \sum_{t\in T}\pi^{E}_{t}E_{t}+\sum_{t\in T} \pi^{R}_{t}R_{t}+\sum_{s\in N_{s}}\omega_{s}\sum_{t\in T}\sum_{i\in N_{p}} \{ U_{s,i,t}-C_{s,i,t}\},
 \vspace{-0.3cm}
\end{align}
where the objective function (4) consists of three terms: (i) the expected revenue from the energy market, (ii) the expected revenue from the reserve market, and (iii) expected surplus of all prosumers in all scenarios. This paper employs a quadratic form for utility functions for prosumer $i$'s power consumption: 
\begin{equation}
    U_{s,i,t}=-\alpha^{U}_{i}(P^{d}_{s,i,t}-P^{D}_{s,i,t})^{2}+\beta^{U}_{i}(P^{d}_{s,i,t}-P^{D}_{s,i,t}),
    \vspace{-0.2cm}
\end{equation}
where $\alpha^{U}_{i}/\beta^{U}_{i}\geq 0$ are utility coefficients of prosumer $i$. And prosumer $i$'s costs include the generation costs of thermal generators and the degradation costs of storage systems\cite{tushar2018peer}.
\begin{align}
 C_{s,i,t}&=C^{g}_{s,i,t}+C^{S}_{s,i,t},\\
 C^{g}_{s,i,t}&=\alpha^{g}_{i} (P^{g}_{s,i,t})^{2}+\beta^{g}_{i} P^{g}_{s,i,t},\\
   C^{S}_{s,i,t}&=\alpha^{S}_{i}| P^{dis}_{s,i,t}-P^{c}_{s,i,t}|+\beta^{S}_{i},
\end{align}
 where $\alpha_{i}/\beta_{i}\geq 0$ are resource-specific cost coefficients.

\subsection{Constraints}
The scenario-based optimal operation model is subject to following constraints:

\emph{1) Limits on the DER's Output:} Equations (9-12) describe the output limit of distributed thermal generators, flexible demands and storage systems. 
 \begin{gather}
  \underline{P}^{g}_{i} \leq P^{g}_{s,i,t} \leq \overline{P}^{g}_{i},\\
 0\leq P^{D}_{s,i,t}\leq \overline{P}^{D}_{i,t},\\
  0\leq P^{dis}_{s,i,t}\leq \overline{P}^{S}_{i},\\
  0\leq P^{c}_{s,i,t}\leq \overline{P}^{S}_{i}.
 \end{gather}

\emph{2) Upward reserve and ramping limits:} The upward reserve contributed by thermal generators within each hour is given by constraint (13), which is coupled with ramping up and down limits described in Eqs. (14) and (15) across consecutive time periods.
\begin{align}
    & 0\leq P^{R}_{s,i,t}\leq min\{10\times MSR_{i},\overline{P}^{g}_{i}-P^{g}_{s,i,t}\},\\
& P^{g}_{s,i,t}-P^{g}_{s,i,t-1}+P^{R}_{s,i,t}\leq RU_{i},\\
& P^{g}_{s,i,t-1}-P^{g}_{s,i,t}+P^{R}_{s,i,t-1}\leq RD_{i}.
\end{align}
Note that (14) means the increase in the output from a previous moment to the output at this moment plus the upward spinning reserve must comply with the upward ramping limit, and vice versa for downward ramping limit.

\emph{3) Constraints of storage systems:} Storage systems cannot charge and discharge at the same time, which is expressed by the following constraints:
\begin{gather}
0\leq P^{c}_{s,i,t} \leq M z_{i,t},\\
0\leq P^{dis}_{s,i,t}\leq M(1-z_{i,t}),
\end{gather}
where $M$ is a large number linked with binary variables, $z_{i,t}$. 

Constraints (18) and (19) establish the energy state of the storage system, ensuring it satisfies the required capacity of storage systems.
\begin{gather}
 Q^{S}_{s,i,t}-Q^{S}_{s,i,t-1}=P^{c}_{s,i,t-1}-P^{dis}_{s,i,t-1},\\
 Q^{0}_{i}\leq Q^{S}_{s,i,t} \leq \overline{Q}_{i}.
\end{gather}

\emph{4) Prosumer's power balance constraint:} Constraint (20) represents the power balance of each prosumer.
\begin{equation}
P^{g}_{s,i,t}+P^{PV}_{s,i,t}+\eta_{i}P^{dis}_{s,i,t}-\frac{1}{\eta_{i}}P^{c}_{s,i,t}-P^{d}_{s,i,t}+P^{D}_{s,i,t}=P^{E}_{s,i,t}+P^{B}_{i,t},
\end{equation}
where the net P2P trading quantity, $P^{B}_{i,t}$, of prosumer $i$ at time $t$ is the $t^{th}$ element in $P^{B}_{i}$. $P^{B}_{i,t}>0$ means that this amount of power is exported, and prosumer $i$ is a P2P seller and vice versa. 

\emph{5) Power flow modelling:} Under the assumption that one node has only one prosumer at most, the nodal active and reactive power injection equations are given by following constraints based on a linear Distflow model from \cite{Linlinear}, $\forall k,i,l\in Nb \subseteq N_{p}, \forall ki\in Nl$
\begin{align}
p_{i,t}&=\sum (p_{ki,t}+\frac{Gs_{ki}}{2}(v_{s,i}+v_{s,k}))-\sum p_{s,jl,t},\notag\\
q_{i,t}& =\sum (q_{ki,t}-\frac{Bs_{ki}}{2} (v_{i}+v_{k}))-\sum q_{il,t},\notag
\end{align}
where the active and reactive power losses are ignored during the linearization. Therefore, we add the terms of power loss $r_{ki}\ell_{s,ki,t}, x_{ki}\ell_{ki,t}$ in the active and reactive power injection equation, which are (21) and (22), respectively.
\begin{align}
p_{s,i,t}&=\sum (p_{s,ki,t}+\frac{Gs_{ki}}{2}(v_{s,i}+v_{s,k})-r_{ki}\ell_{s,ki,t})-\sum p_{s,il,t}\notag\\
&=-P^{E}_{s,i,t}-P^{B}_{s,i,t},\\
q_{i,t}& =\sum (q_{ki,t}-\frac{Bs_{ki}}{2} (v_{i}+v_{k})-x_{ki}\ell_{ki,t})-\sum q_{il,t}=-q^{d}_{i,t}.
\end{align}

Then the squared current magnitude in the power loss terms are expressed in (23) based on the second order cone programming:
\begin{equation}
   \ell_{s,ki,t}\geq \frac{p^{2}_{s,ki,t}+q^{2}_{ki,t}}{v_{s,i,t}}.
\end{equation}
where $\ell_{0i}\:=0$ when bus $i$ is the substation bus.

The voltage drop equation is given by
\begin{align}
v_{s,i,t}=\gamma^{*}_{ki}v_{s,k,t}-2(&r_{ki}p_{s,ki,t}+x_{ki}q_{ki,t})+(r^{2}_{ki}+x^{2}_{ki})\ell_{s,ki,t}.
\end{align}
Moreover, branch flow limits and nodal voltage magnitude limits are given by constraints (25) and (26), respectively.
\begin{gather}
     S_{s,il,t}\leq \overline{S}_{il},\\
 \underline{V}^{2}_{i}\leq v_{s,i,t}\leq \overline{V}^{2}_{i}.
 \vspace{-0.3cm}
\end{gather}

\emph{6) Energy and reserve provided in the wholesale market:} The energy that the DSO submits in the wholesale market is given in (27). Note that the power provided by DSO in the wholesale market is determined at the substation node, by subtracting the total power loss from the sum of all prosumers' power output for energy market at different nodes. Equation (28) means that the upward reserve is contributed by all prosumers. 
 \begin{gather}
 E_{t}= \sum_{i\in N_{p}}P^{E}_{s,i,t}-\sum_{kj\in N_{l}}r_{ki}\ell_{s,ki,t},\\
  R_{t}=\sum_{i\in N_{p}}P^{R}_{s,i,t}.
  \vspace{-0.5cm}
\end{gather}
Moreover, the supply or demand for the energy and reserve market must satisfy physical transmission constraints at the substation node, denoted by (29) and (30).
\begin{gather}
     |E_{t}|\leq \overline{E}_{ex},\\
 |E_{t}+R_{t}|\leq \overline{E}_{ex}.
 \vspace{-0.3cm}
\end{gather}

Then we can summarize the proposed model for DSO's optimal operation:
\vspace{-0.1cm}
\begin{equation}
\begin{array}{l}
{\rm{\textbf{P1}}}\;\;\;\;\;\;\;\;\;\;\;\;\;\;\;\;\;\;\;\;\,\mathop {\max }\limits_{\mathcal{O}} :{\kern 1pt} \,\,(4)\\
\;\;\;\;\;\;\;\;\;\;\;\;\;\;\;\;\;\;\;\;\;\;\;{\kern 1pt} \;\;{\rm{s}}{\rm{.t}}\;\;\;\;\;\;\;{\kern 1pt}(9 - 30){\rm{, }}\forall s \in {N_s},\forall i \in {N_p},\forall t \in T, \notag
\end{array}
\vspace{-0.25cm}
\end{equation}
where decision variables $\mathcal{O}$ include the energy and reserve provided by DSO in the wholesale market, $E_{t},R_{t}$, their corresponding contributions from internal prosumers, $P^{E}_{s,i,t}, P^{R}_{s,i,t}$, and optimal dispatching of DERs, $P^{g}_{s,i,t}, P^{c}_{s,i,t}, P^{dis}_{s,i,t}, P^{D}_{s,i,t}, Q^{S}_{s,i,t}$. If $E_{t}<0$, DSO buys from the energy market and $\pi^{E}_{t}E_{t}$ is the expected purchase cost. 

 The DSO solves the optimal dispatching problem \textbf{P1} including all generated scenarios and time intervals together. We skip the solution process here and \textbf{P1} can be solved easily with off-the-shelf solvers, e.g. gurobi. Besides, we use an exact relaxation method in \cite{lizhengshuo} to relax the non-convex problem \textbf{P1}, mainly constraints (16-17), to a convex form under two sufficient conditions, which are proved to hold in most cases. 
 
 As shown in (20) and (21), the P2P trading quantity, which constitutes a portion of the net power output/input of prosumers, may affect their power supply or demand in the energy market. Therefore, the DSO must understand the extent to which P2P trading would affect its optimal dispatch.

\section{Impact of P2P Trading to DSO's Optimal Operation}
For the energy contract P2P trading, we have the following Theorem:

\begin{theorem}
Consider the relaxed DSO's optimal dispatch problem \textbf{P1}, define the set of all decision variables except $P^{E}$ and $E$ as $\mathcal{O}'$ and their optimal values solved at $P^{B}=0$ as $\mathcal{O}^{'*}_{0}$. Then $\forall |P^{B}|\neq 0 $, there exist a set of $P^{E*}:=P^{E,0}-P^{B}$ and a set of $E^{*}=E^{0}-\sum_{i\in N_{p}}P^{B}$ such that $\mathcal{O}^{*}=\{\mathcal{O}^{'*}_{0}, P^{E*}, E^{*}\}$ is optimal for \textbf{P1}. In addition, if there is $\sum_{i\in N_{p}}P^{B}=0$, then $E^{*}=E^{0}$. \footnote{ For simplicity, we include subscripts of $P^{B}_{i,t},P^{E}_{s,i,t}$ and $E_{t}$ in \textbf{P1} as $P^{B},P^{E}$ and $E$, whose optimal value solved at $P^{B}=0$ is denoted by $P^{E,0}$ and $E^{0}$.} 
\end{theorem}

\noindent \textbf{Proof:} 
Denote four constraints, Eqs. (20), (21), (27) and (28), that are correlated to P2P trading quantity $P^{B}$ as $H1, H2, H3$ and $H4$. Then we can write the compact form of model \textbf{P1} by summarizing all other inequality and equality constraints:
    \begin{equation}
\left\{ \begin{array}{l}
\min \; - S(O),\\
{\rm{s}}{\rm{.t}}{\rm{.}}\;\;\;{f_n}(O) \le 0,\;\;\;\;\;\;n = 1,{\kern 1pt} {\kern 1pt}  \cdots ,N,\\
\,\,\quad \;\;{\kern 2pt} {a_m}O = {b_m},\;\;\;\;\,{\kern 1pt} m = 1,{\kern 1pt} {\kern 1pt}  \cdots ,M,\\
H1,\;H2,\;H3,\;H4,\,\;\;\forall s \in {N_s},\forall i \in {N_p},\forall t \in T.
\end{array} \right.
    \end{equation}

Denote multipliers of $H1, H2$ that are dependent on scenarios as $\lambda _{s,i,t}^P,\lambda _{s,i,t}^I$ and that of $H3$ and $H4$ which do not include scenarios as $\lambda _t^E$ and $\lambda _t^R$, respectively. Subsequently, multipliers of other inequality constraints and equality constraints are denoted by ${\mu _n}$ and ${\lambda _m}$, respectively. As a result, the Lagrangian of \textbf{P1} can be written as
\begin{equation}
\begin{array}{l}
L(O,{\mu _n},{\lambda _m},\lambda _{s,i,t}^P,\lambda _{s,i,t}^I,\lambda _t^E,\lambda _t^R) = \\
 - S(O) + \sum\limits_{n = 1}^N {{\mu _n}{f_n}(O) + } \sum\limits_{m = 1}^m {{\lambda _m}({a_m}O - {b_m}) + } \sum\limits_{s \in {N_s}}^{} {\sum\limits_{i \in {N_p}}^{} {\sum\limits_{t \in T}^{} {\lambda _{s,i,t}^P} } } \\
H1 + \sum\limits_{s \in {N_s}}^{} {\sum\limits_{i \in {N_p}}^{} {\sum\limits_{t \in T}^{} {\lambda _{s,i,t}^IH2} } }  + \sum\limits_{s \in {N_s}}^{} {\sum\limits_{t \in T}^{} {\lambda _t^E} } H3 + \sum\limits_{s \in {N_s}}^{} {\sum\limits_{t \in T}^{} {\lambda _t^RH4} } .
\end{array}
\end{equation}
Then, the corresponding Karush-Kuhn-Tucker (KKT) conditions are:
\begin{equation}
\left\{ \begin{array}{l}
{\nabla _O}L(O,{\mu _n},{\lambda _m},\lambda _{s,i,t}^P,\lambda _{s,i,t}^I,\lambda _t^E,\lambda _t^R) = 0,\\
H1,\;H2,\;H3,\;H4,\,\;\;\forall s \in {N_s},\forall i \in {N_p},\forall t \in T,\\
{a_m}O - {b_m} = 0,\quad \;\;\;\;\;m = 1, \cdots ,M,\\
{f_n}(O) \le 0\,;\;\;\;\;\;\,\;\,\,\,\,\,\;\;\;{\kern 1pt} \,n = 1, \cdots ,N,\\
{\mu _n}{f_n}(O) = 0,{\kern 1pt} \,\,\;\;\;\;\,\;\;\;\,\,{\kern 1pt} {\kern 1pt} n = 1, \cdots ,N,\\
{\mu _n} \ge 0,\,\,\;\quad \quad \,\;\;\,\;\;\;\;\;\,\,{\kern 1.3pt} {\kern 1pt} n = 1, \cdots ,N.
\end{array} \right.
\end{equation}

Denote $\mathcal{O}^{*}=\{\mathcal{O'}^{*}, P^{E*}_{s,i,t}\}$ as the optimal solution to model \textbf{P1} that satisfies the above KKT conditions, along with corresponding multipliers $\mathcal{M}^*=\{{\mu^* _n},{\lambda^* _m},\lambda _{s,i,t}^{P*},\lambda _{s,i,t}^{I*},\lambda _t^{E*},\lambda _t^{R*}\}$. When prosumers do not have P2P transactions, $P^{B}_{i,t}=0$, we can find $\mathcal{O}^{'*}_{0}$ and $P^{E,0}_{s,i,t}$ with $\mathcal{M}^{*}_{0}$ satisfying the above KKT conditions (33), where conditions $H1,H2,H3$ and $H4$, $\forall s\in N_{s}, \forall i\in N_{p}, \forall t\in T$, are expressed as:
\begin{equation}
    \left\{ \begin{array}{l}
P_{s,i,t}^{{g^*},0} + P_{s,i,t}^{PV}+\eta_{i}P_{s,i,t}^{{dis^*},0}- \frac{P_{s,i,t}^{{c^*},0}}{\eta_{i}} - P_{s,i,t}^d + p_{s,i,t}^{{D^*},0} - P_{s,i,t}^{{E},0} = 0,\\
P_{s,i,t}^0 + P_{s,i,t}^{{E},0} = 0,\\
E_t^{0} - \sum\limits_{i = 1}^{{N_p}} {P_{s,i,t}^{{E},0}}  - \sum\limits_{ki = 1}^{{N_l}} {{r_{ki}}\ell_{_{s,ki,t}}^0}  = 0,\\
R_t^{0} - \sum\limits_{i = 1}^{{N_p}} {P_{s,i,t}^{{R^*},0} = 0.\;} 
\end{array} \right.
\end{equation}

When P2P trading happens between prosumers, $P^{B}_{i,t}\neq 0$. $\mathcal{O}^{'*}_{0}$ and $\mathcal{M}^{*}_{0}$ still meet KKT conditions (33) except $H1,\dots,H4$ as P2P trading may affect these equality constraints. Subsequently, input $\mathcal{O}^{'*}_{0}, P^{E*}_{s,i,t}=P_{s,i,t}^{{E},0}-P^{B}_{i,t}$ and $E^{*}_{t}=E^{0}_{t}-\sum_{i\in N_{p}}P^{B}_{i,t}$ into $H1, H2, H3$ and $H4$ in particular $\forall s\in N_{s}, \forall i\in N_{p}, \forall t\in T$:

 For constraint $H1$:
    \begin{align}
   H1&=P_{s,i,t}^{{g^*},0} + P_{s,i,t}^{PV}  + \eta_{i}P_{s,i,t}^{{dis^*},0}- \frac{P_{s,i,t}^{{c^*},0}}{\eta_{i}} - P_{s,i,t}^d + P_{s,i,t}^{{D^*},0} - P_{s,i,t}^{{E^*}} \notag\\
&- P_{i,t}^{B} =P_{s,i,t}^{{g^*},0} + P_{s,i,t}^{PV}  + \eta_{i}P_{s,i,t}^{{dis^*},0}- \frac{P_{s,i,t}^{{c^*},0}}{\eta_{i}} - P_{s,i,t}^d + P_{s,i,t}^{{D^*},0} \notag\\&- P_{s,i,t}^{E,0} + P_{i,t}^{B} - P_{i,t}^{B} =
 P_{s,i,t}^{{g^*},0} + P_{s,i,t}^{PV}  + \eta_{i}P_{s,i,t}^{{dis^*},0}- \frac{P_{s,i,t}^{{c^*},0}}{\eta_{i}}\notag\\&- P_{s,i,t}^d + P_{s,i,t}^{{D^*},0} - P_{s,i,t}^{E,0} = 0.
 \end{align}

 For constraint $H2$:   
 \begin{equation}
H2=p_{s,i,t}^0 + P_{s,i,t}^{{E^*}} + P_{i,t}^{B}  = p_{s,i,t}^0 + P_{s,i,t}^{E,0} = 0.
 \end{equation}

 For constraint $H3$: 
 \begin{align}
H3&= E_t^* - \sum\limits_{i \in {N_p}} {P_{s,i,t}^{{E^*}}}  - \sum\limits_{ki \in {N_l}} {{r_{ki}}l_{_{s,ki,t}}^0}  \notag\\&= E_t^* - \sum\limits_{i \in {N_p}} {(P_{s,i,t}^{{E},0} - } P_{i,t}^{B})- \sum\limits_{ki\in {N_l}} {{r_{ki}}l_{_{s,ki,t}}^0}  \notag\\&= E_t^{0} - \sum\limits_{i \in {N_p}} {P_{i,t}^{B}}  - \sum\limits_{i \in {N_p}} {P_{s,i,t}^{{E},0} + } \sum\limits_{i \in {N_p}} {P_{i,t}^{B}} - \sum\limits_{ki \in {N_l}}
{{r_{ki}}l_{_{s,ki,t}}^0}\notag\\&= E_t^{0} - \sum\limits_{i \in {N_p}} {P_{s,i,t}^{E,0}}  - \sum\limits_{ki \in {N_l}} {{r_{ki}}l_{_{s,ki,t}}^0}  = 0,
 \end{align}
 
  For constraint $H4$: 
 \begin{equation}
     H4=R_t^{0} - \sum\limits_{i = 1}^{{N_p}} {P_{s,i,t}^{{R^*},0} = 0.\;} 
 \end{equation}

Note that Eqs. (35-38) when $P^{B}_{i,t}\neq 0$ are exactly the same as (34) when $P^{B}_{i,t}= 0$, which means that $P^{E*}_{s,i,t}$ and $E_t^*$ are bounded to meet the KKT conditions. Therefore, $\mathcal{O}^{'*}_{0}$, a set of $P^{E*}_{s,i,t}$ and a set of $ E_t^*$ are the optimal solution of the proposed model \textbf{P1} under different P2P trading parameters.$\hfill\square$

\noindent \textbf{Remark 1.} As a special case of energy contract P2P trading where real-time trading power matches, the power contract P2P trading restricts $\sum\limits_{i \in {N_p}} P_{i,t}^B = 0$, therefore Theorem 1 shows that the power contract P2P trading has no impact on DSO's optimal dispatch of DERs, energy and reserve provided to the wholesale market, and surplus, as their optimal solutions and values are the same as that without P2P trading. Therefore, the DSO can safely disregard the power contract P2P trading between internal prosumers in its optimal operation.

\noindent \textbf{Remark 2.} The supply or demand for energy in the wholesale market and total surplus are affected by various P2P trading quantities, because the real-time trading powers of buyer and seller in each energy contract P2P transaction may be different, $\sum\limits_{i \in {N_p}} P_{i,t}^B \neq 0$. Moreover, if the common part between buyer's output and seller's demand takes a higher portion, the extent of the impact from P2P trading will decrease as the real-time trading power mismatches are also reduced. 

\noindent \textbf{Remark 3.} The surplus of all prosumers is contributed together by net P2P trading power, $P_{i,t}^B$ and power for energy market, $P_{i,t}^E$, whose summation is net nodal power injection. As the net nodal power injection remains the same, prosumers can ignore the corresponding power loss and line congestion when deciding P2P trading under the framework in this paper, while DSO can also safely disregard the change in these two terms when P2P trading changes.

\section{Case Study}
\subsection{Settings}
Simulations have been done on one simple, 2-bus system, and a real network, United Kingdom Generic Distribution System (UKGDS) 95-bus system. Totally 1000 scenarios are generated by Monte Carlo method, in which uncertainties follow independent normal distributions:
\begin{equation}
    P^{PV}_{s,i,t}\sim N(P^{PV}_{i,t},\sigma^{2}_{PV}),
\end{equation}
\begin{equation}
    P^{d}_{s,i,t}\sim N(P^{d}_{i,t},\sigma^{2}_{d}),
\end{equation}
where the mean value $P^{PV}_{i,t}, P^{d}_{i,t}$ are predicted values, and corresponding real-time predicted errors are expressed by standard deviations, $\sigma$.  

\begin{figure}[ht!]
  \centering
  \includegraphics[scale=0.4]{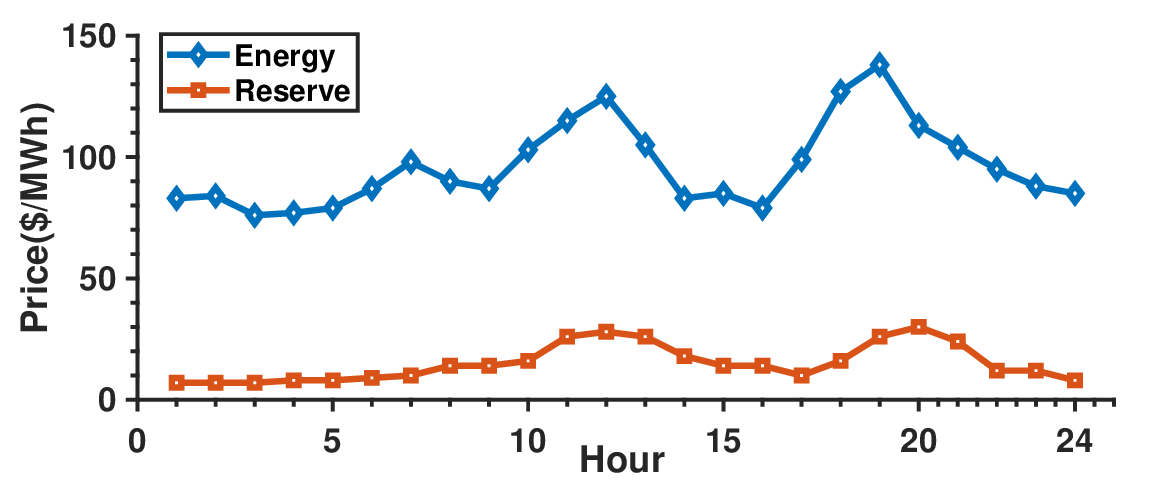}
  \captionsetup{singlelinecheck=off}
  \caption{Price setting}
  \vspace{-0.5cm}
\end{figure}

The predicted market prices, as shown in Fig. 3, used in these two test systems are set in accordance to the average price of the energy and reserve market organised by CAISO in 2022\cite{CAISO_reg_reserve_req_ratio}.  

The P2P trading ratio is defined as $P^{B}_{i,t}/(min\{P^{d}_{s,i,t}\}-\overline{P}^{D}_{i,t})$, which is the net P2P trading quantity divided by the smallest load of the P2P buyer over all scenarios. This ratio represents to what extent the P2P buyer meets its own demand through P2P transactions. If the P2P trading ratio is 0, it means that there is no P2P trading. When the trading ratio increases to 1, the smallest P2P buyer's demand among all scenarios is satisfied through P2P trading. We further explored the situations when P2P trading ratio is higher than 1, means that the P2P buyer can also export its extra power from P2P trading to DSO after fulfilling its own demand.

Three economic indices are introduced to further demonstrate the impact of P2P trading:

  \begin{itemize}
      \item Total surplus: is the objective function in the DSO's optimal operation.
      \item Net surplus of prosumers and P2P trading: is the summation of the net surplus and the utility of prosumers consuming P2P trading quantities. The net surplus is the total surplus deducing the utility of all prosumers. Net surplus of prosumers and P2P trading can be calculated as:
      \begin{equation}
          \sum_{t\in T}\pi^{E}_{t}E_{t}+\sum_{t\in T} \pi^{R}_{t}R_{t}+\sum_{t\in T}\sum_{i\in N_{p}} \{ U_{i,t}(P^{B}_{i,t})-C_{i,t}\}
      \end{equation}
      $\forall s \in N_{s}$. This index provides a comprehensive assessment of the combined surplus of the DSO and P2P trading activities.
    \item Incremental improvement of P2P Trading: quantifies the additional benefit from the utilization of P2P trading quantities compared to the case that such quantities are entirely utilized by the DSO, eliminating P2P trading. Incremental improvement is calculated as the difference between the net surplus of prosumers and P2P trading and the net surplus of prosumers without P2P trading.
  \end{itemize}

\subsection{2-bus System}
Two prosumers are located at opposite ends of the branch, as shown in Fig. 4. The detailed curve of forecasted PV output and load power is shown in Fig. 5. The standard deviation for PV output and load power is 0.8 MW and 0.3 MW, respectively. The rest setting of prosumers' DERs are given in Table I. 
\begin{table}[ht!]
\centering
   \caption{Parameter setting of prosumers' DERs (MW)} 
   \label{Non-base}
   \begin{tabular}{ccc}
    \hline
    \hline
Parameter & Prosumer 1 & Prosumer 2\\
   \hline
        $\underline{P}^{g}$,$\overline{P}^{g}$ & 0.25,5 & 0,0\\
        $MSR,RU/RD$ & 0.05,1.5 & 0,0\\
        $\overline{P}^{D}$ & 0 & 0.25\\
        $\overline{P}^{S}$ & 0.25 & 0.25\\
        $\eta$ & 0.8 & 0.9\\
        $Q^{0},\overline{Q}$ & 0,1.25 MWh & 0,1.25 MWh\\
        \hline
    \hline
   \end{tabular}
     \vspace{-0.3cm}
  \end{table}

\begin{figure}[ht!]
  \centering
  \includegraphics[scale=0.5]{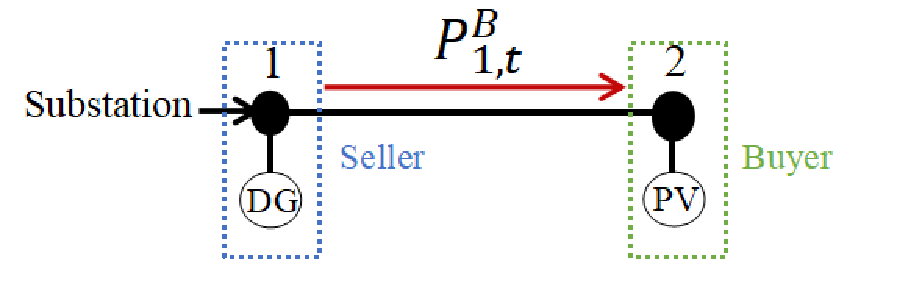}
  \captionsetup{singlelinecheck=off}
  \caption{ A 2-bus system}
  \vspace{-0.3cm}
\end{figure}
  \begin{figure}[ht!]
  \centering
  \includegraphics[scale=0.43]{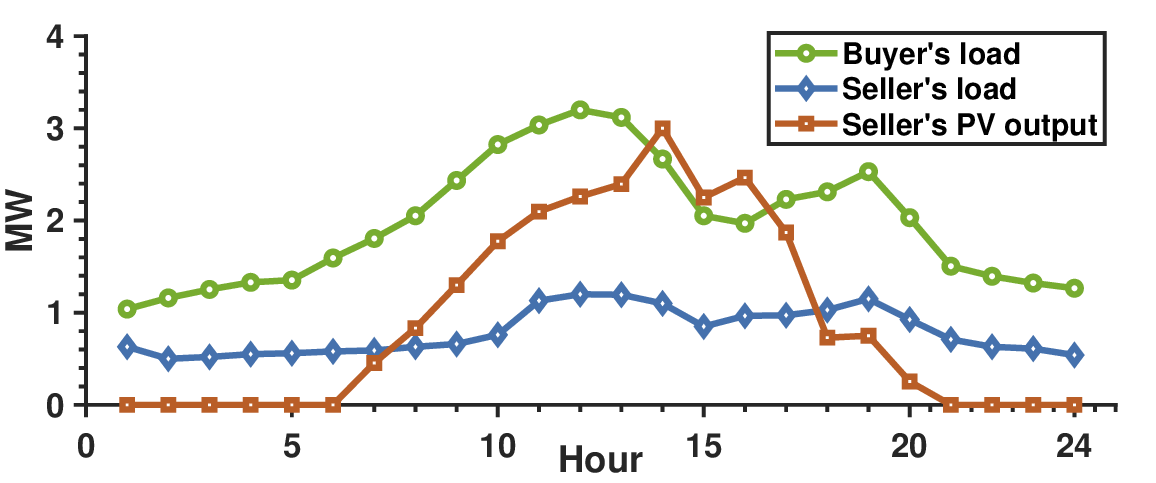}
  \captionsetup{singlelinecheck=off}
  \caption{ Prosumers' PV output and load power}
  \vspace{-0.5cm}
\end{figure}In this case, prosumer 1 is designated as the seller, while prosumer 2 takes on the role of the buyer. The maximum P2P trading quantity from prosumer 1 to prosumer 2 is subject to the limitation imposed by the latter's demand, taking into account the deduction of the maximum load curtailment. The substation node is subjected to an exchanging limit of 5 MW. 

We tested the DSO's optimal operation model in the 2-bus system on a 24-hour scale. The simulation focuses on the change in energy and reserve provided by DSO in the wholesale market and the total surplus of all prosumers by gradually lifting the P2P trading ratio. 

\begin{figure*}[htbp]
\centering
\subfigure{
\begin{minipage}[t]{0.32\linewidth}
\centering
\includegraphics[width=\linewidth]{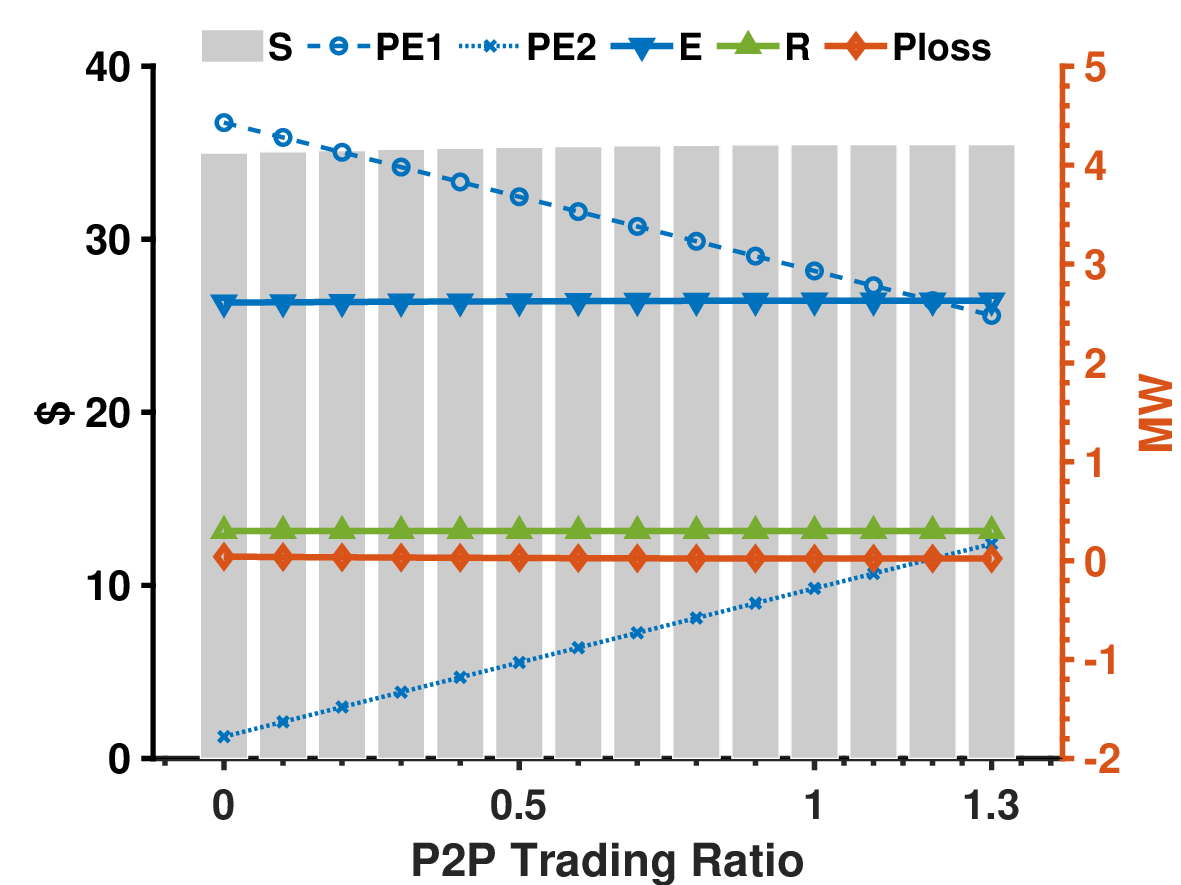} 
\label{(a)}
\end{minipage}}
\subfigure{\begin{minipage}[t]{0.32\linewidth}
\centering
    \includegraphics[width=\linewidth]{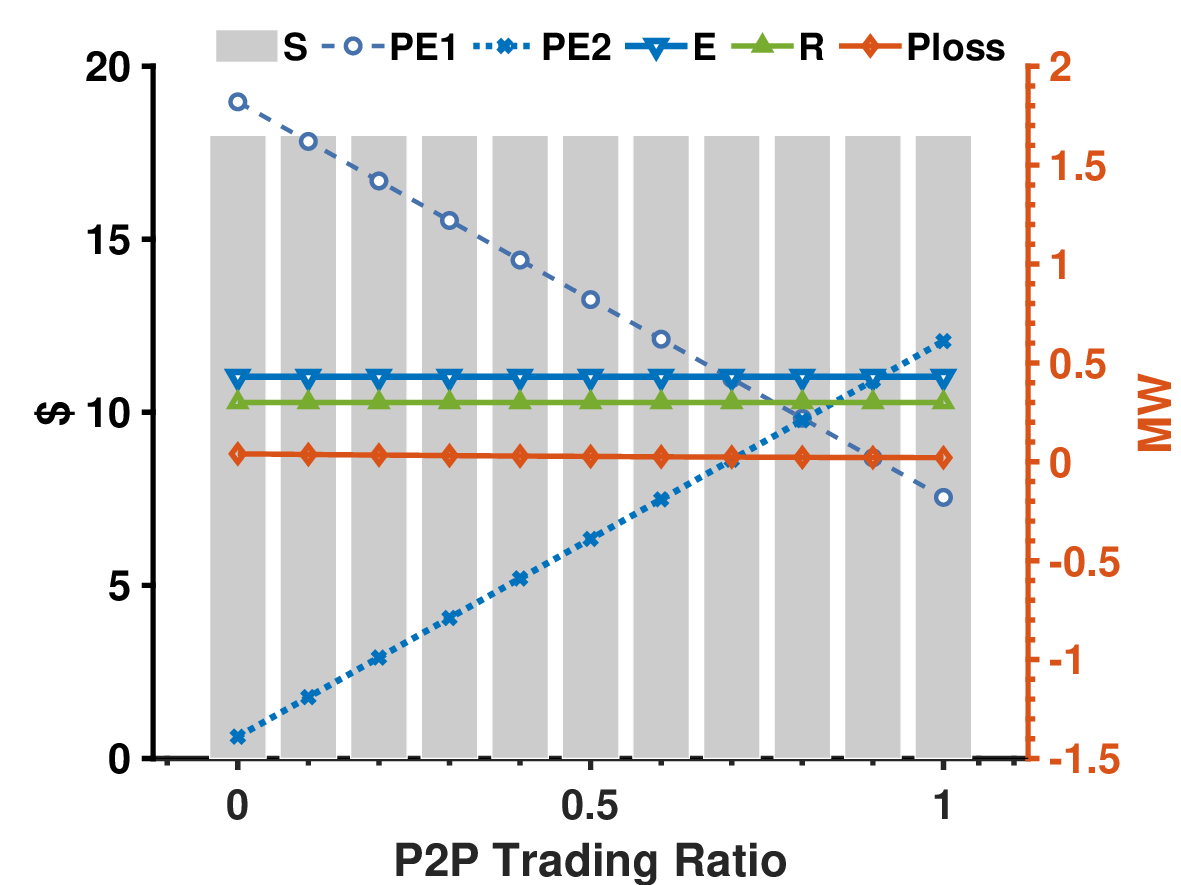}
    \label{(b)}
\end{minipage}}
\subfigure{\begin{minipage}[t]{0.32\linewidth}
\centering
    \includegraphics[width=\linewidth]{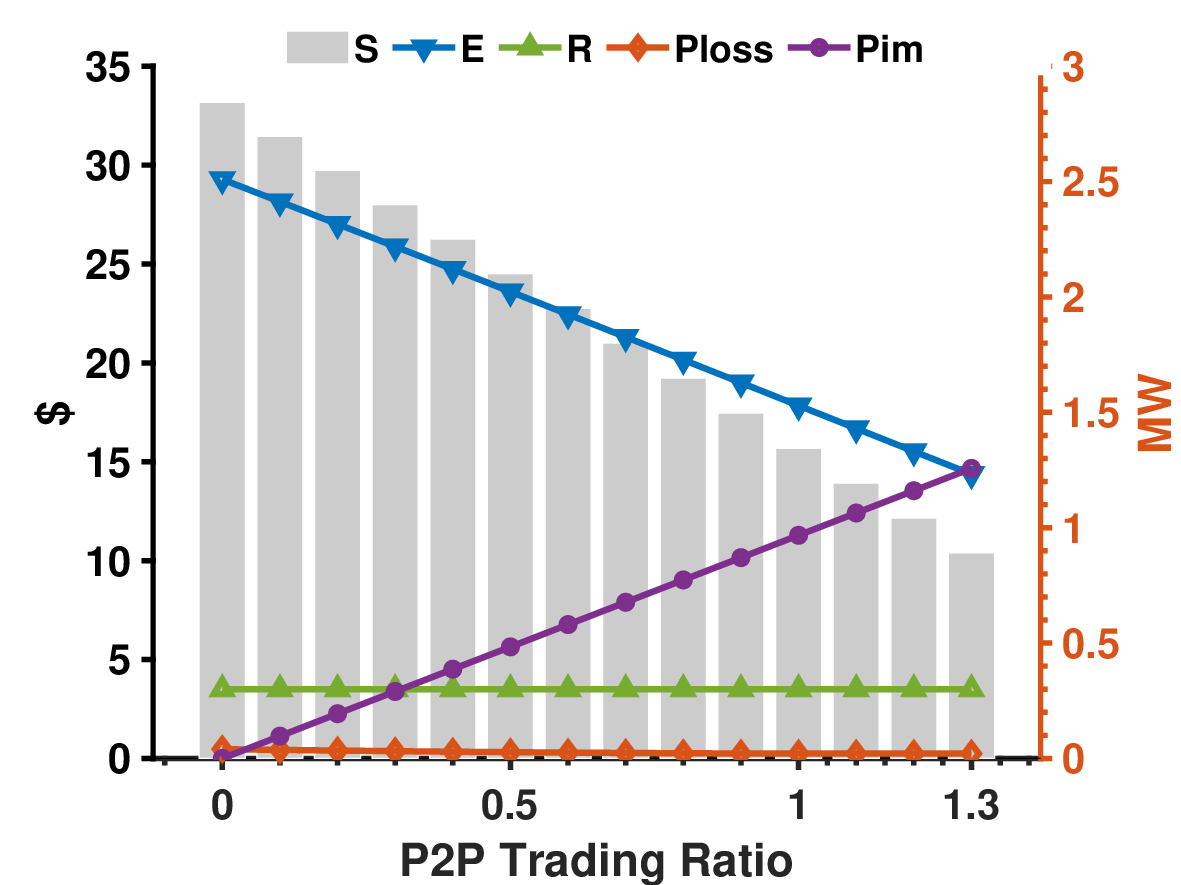}
    \label{(c)}
\end{minipage}}
\centering
  \captionsetup{singlelinecheck=off}
\caption{Simulation results under different P2P trading ratios in 2-bus system: (a) Power contract P2P trading (hour 11), (b) Power contract P2P trading (hour 18) and (c) Energy contract P2P trading (hour 12)}
\vspace{-0.5cm}
\end{figure*}

As for power contract P2P trading, the simulation results at hour 11 are presented in Fig. 6 (a), demonstrating the impact of P2P trading. The power output of P2P seller to the energy market, $P^{E}_{1}$, decreases as P2P trading ratio increases, while that of P2P buyer increases, $P^{E}_{2}$, at the same rate. Note that when the P2P trading ratio is higher than 1, prosumer 2 also submits $P^{E}_{2}$ to DSO to sell in the wholesale market after meeting its own demand through P2P trading. Therefore, the sum of these two variables minus system power loss, representing the energy provided by DSO in the wholesale market is constant when P2P trading ratio changes. The simulation results echo with Theorem 1.

Note that $P^{E}_{1}$ may fall below zero when $P^{B}_{1}$ increases (Fig. 6 (b)) at hour 18, indicating that the P2P seller needs to purchase from the energy market through DSO to fulfill its P2P trading requirements. The reason why prosumer 1 tends to purchase from the market instead of generating more power from its DERs is that the power output of DERs is either at the optimal output point or the upper limit when P2P trading ratio is 0, and therefore prosumer 1 would not generate more for its P2P transaction adjustments.
\begin{table}[ht!]
\centering
   \caption{Economic Indices Under Different Power Contract P2P Trading Ratios} 
   \label{Non-base}
   \begin{tabular}{ccccc}
    \hline
    \hline
    \specialrule{0em}{1pt}{1pt}
\multirow{2}{*}{P2P Ratio} & \multirow{2}{*}{Total Surplus (\$)} & \multirow{2}{*}{Net Surplus (\$)} & Incremental \\ & & & Improvement(\$)  \\
    \specialrule{0em}{1pt}{1pt}
   \hline
   \specialrule{0em}{1pt}{1pt}
0/10 & 34.32 & 21.56 & 0\\
2/10 & 34.32 & 48.47  & 26.91\\
4/10 & 34.32 & 63.75  & 42.19\\
6/10 & 34.32 & 82.12  & 60.56\\
8/10 & 34.32 & 103.06   & 81.50\\
10/10 & 34.32 & 189.67  & 168.11\\
\specialrule{0em}{1pt}{1pt}
        \hline
    \hline
   \end{tabular}
  \end{table}
\begin{table}[ht!]
\centering
   \caption{Economic Indices Under Different Energy Contract P2P Trading Ratios} 
   \label{Non-base}
   \begin{tabular}{ccccc}
    \hline
    \hline
    \specialrule{0em}{1pt}{1pt}
\multirow{2}{*}{P2P Ratio} & \multirow{2}{*}{Total Surplus (\$)} & \multirow{2}{*}{Net Surplus (\$)} & Incremental \\ & & & Improvement(\$)  \\
    \specialrule{0em}{1pt}{1pt}
   \hline
   \specialrule{0em}{1pt}{1pt}
0/10 & 34.32 & 21.56 & 0\\
2/10 & 30.28 & 28.20  & 6.64\\
4/10 & 26.24 & 36.84  & 15.28\\
6/10 & 22.21 & 45.48  & 23.92\\
8/10 & 18.17 & 64.12   & 42.56\\
10/10 & 14.13 & 82.76  & 61.20\\
\specialrule{0em}{1pt}{1pt}
        \hline
    \hline
   \end{tabular}
   \vspace{-0.3cm}
  \end{table}
  
With respect to the energy contract P2P trading, the results are shown in Fig. 6 (c). There exists the real-time trading power imbalance between P2P buyer and seller, $\sum_{i\in N_{p}}P^{B}_{i,t}\neq 0$. When the P2P trading ratio increases, the total P2P trading power imbalance of all P2P transactions, $Pim$, also increases. Therefore, the energy provided by DSO to the wholesale market decreases according to changing P2P trading quantities, as well as the total surplus of all prosumers.

Table II and III give the changes of economic indices under different P2P trading ratios. As discussed, power contract P2P trading does not affect the total surplus. However, an increase in P2P trading quantities leads to a definite enhancement in the net surplus of prosumers and P2P trading, given that the utility of prosumers consuming P2P trading quantities is an increasing function of the trading quantity. The incremental improvement reflects the disparity in prosumers' utilization for P2P trading. Since the net surplus of prosumers remains unaffected by the presence or absence of P2P trading, the incremental improvement is lifted with increasing P2P trading ratios. As for the energy contract P2P trading, the increasing P2P trading ratio decreases the total surplus because the DSO needs to compensate the real-time P2P trading power imbalance through adjusting the submitted energy from prosumers in the market participation. As for the remaining indices, they follow the same tendency with that of power contract P2P trading. \begin{figure}[ht!]
  \centering
  \includegraphics[scale=0.5]{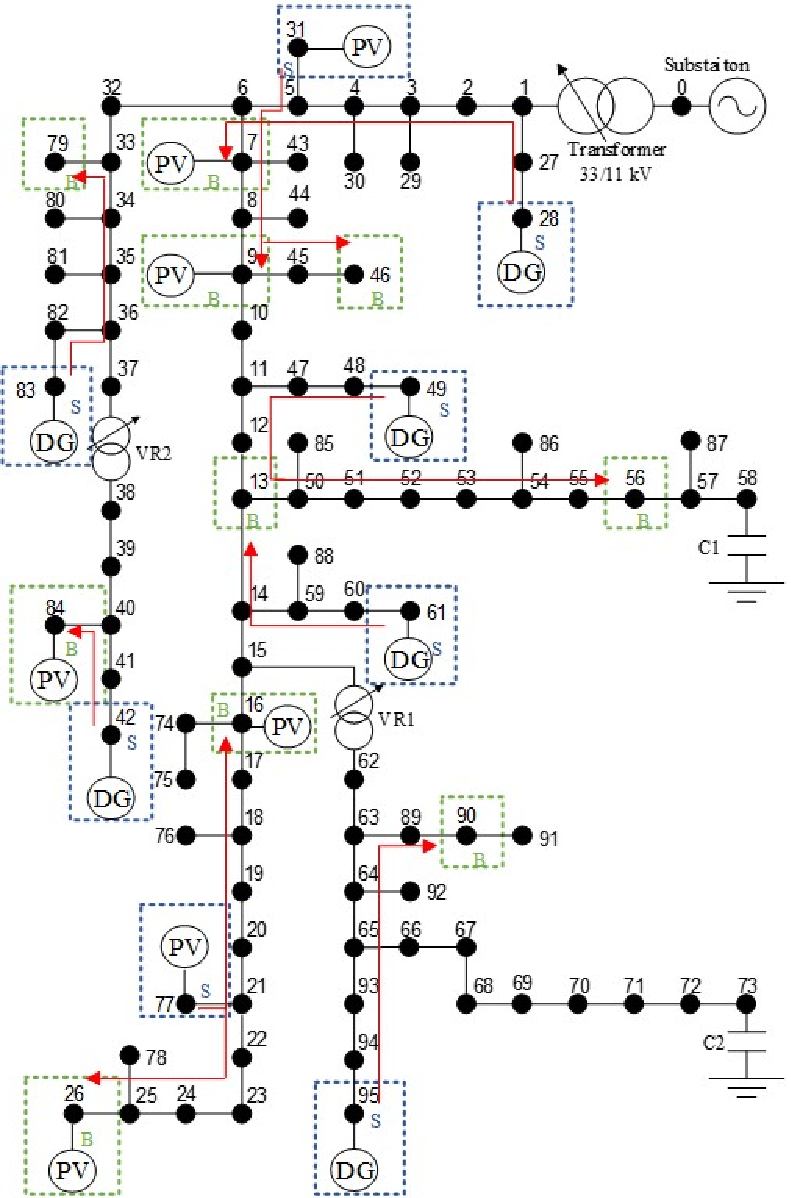}
    \captionsetup{singlelinecheck=off}
  \caption{System setting of UKGDS 95-bus system}
  \vspace{-0.5cm}
\end{figure}The reason why their values increase at a relatively lower rate when compared to power contract P2P trading is that the calculation of net surplus of prosumers and P2P trading and incremental improvement needs to count on the decreasing revenue from the energy market when P2P trading ratio increases. Note that the energy contract P2P trading will decrease the total surplus and increase the utility of the prosumers engaged in P2P trading because they are more willing to meet their demand through P2P trading.

\subsection{United Kingdom Generic Distribution 95-bus System}
\begin{figure*}[htbp]
\centering
\subfigure{
\begin{minipage}[t]{0.32\linewidth}
\centering
\includegraphics[width=\linewidth]{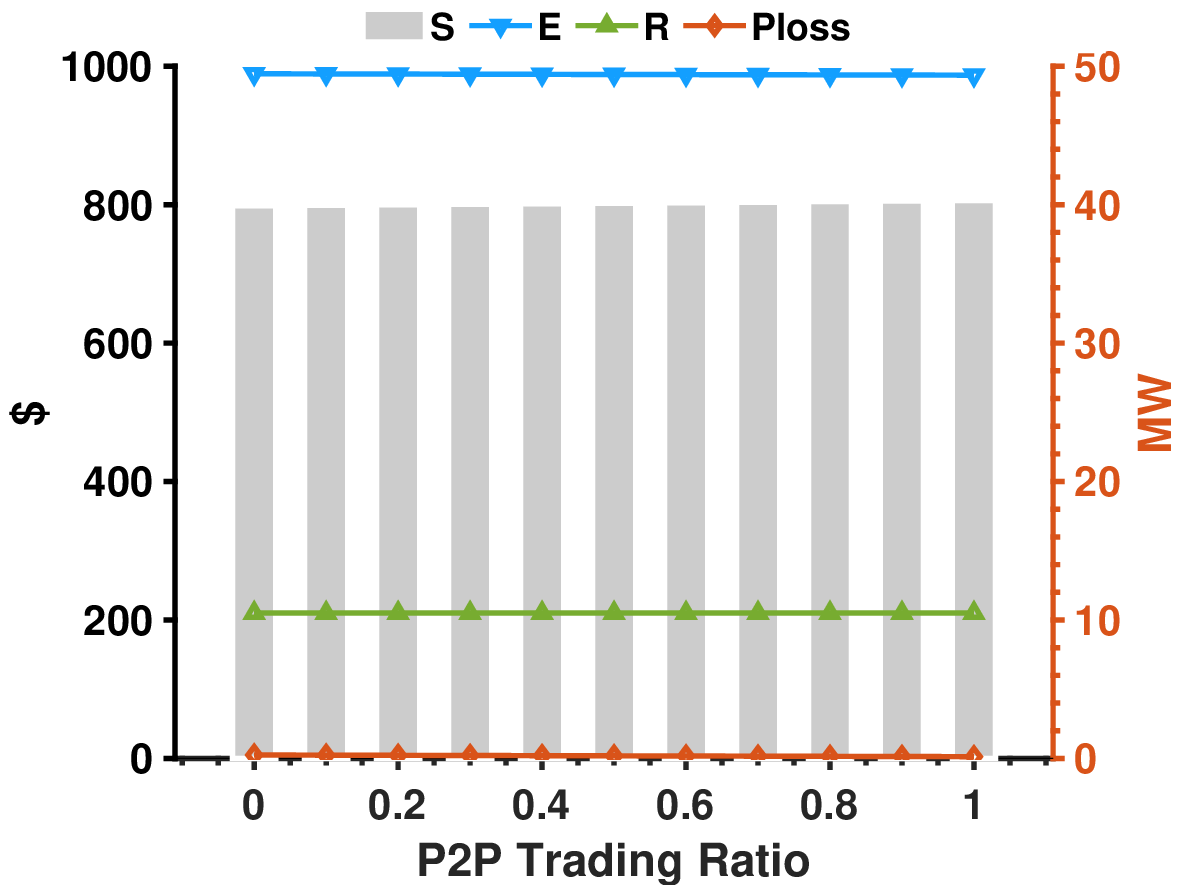} 
\label{(a)}
\end{minipage}}
\subfigure{\begin{minipage}[t]{0.32\linewidth}
\centering
    \includegraphics[width=\linewidth]{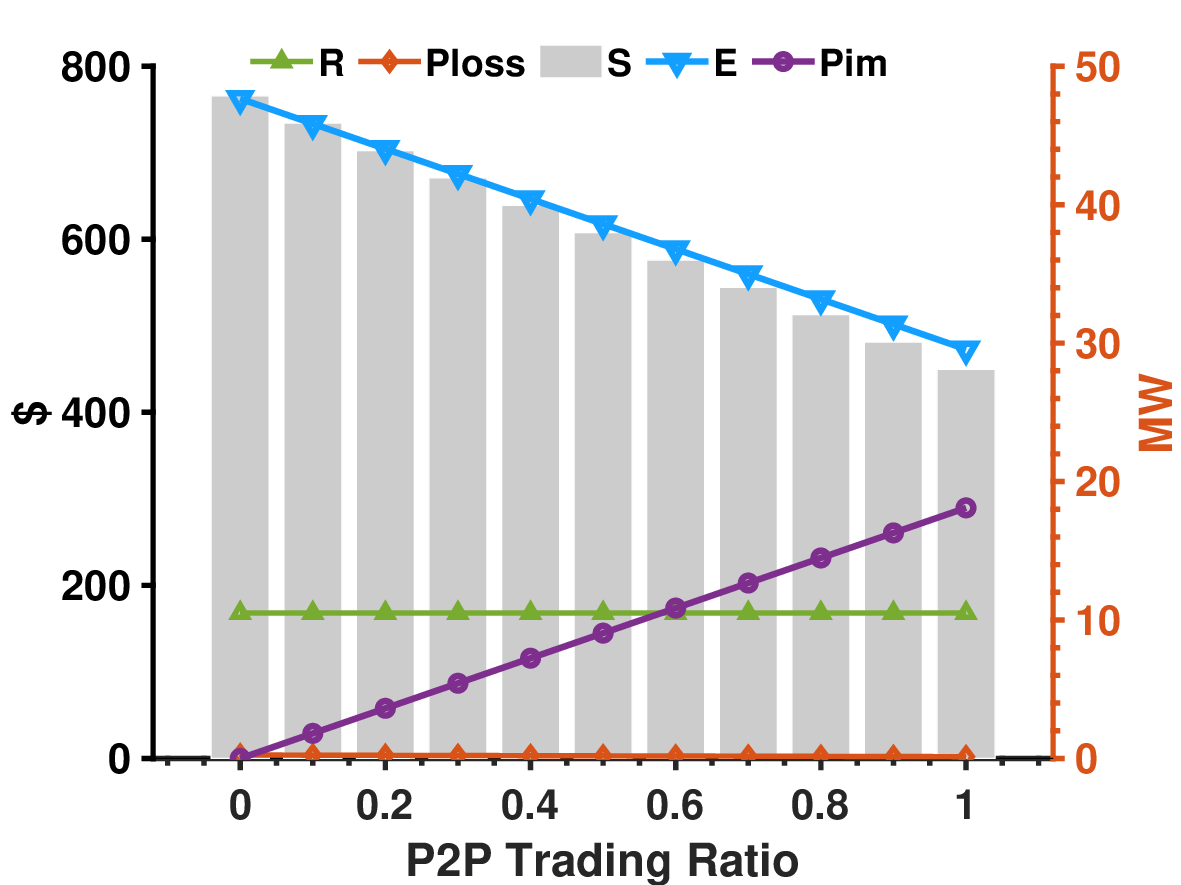}
    \label{(b)}
\end{minipage}}
\subfigure{\begin{minipage}[t]{0.32\linewidth}
\centering
    \includegraphics[width=\linewidth]{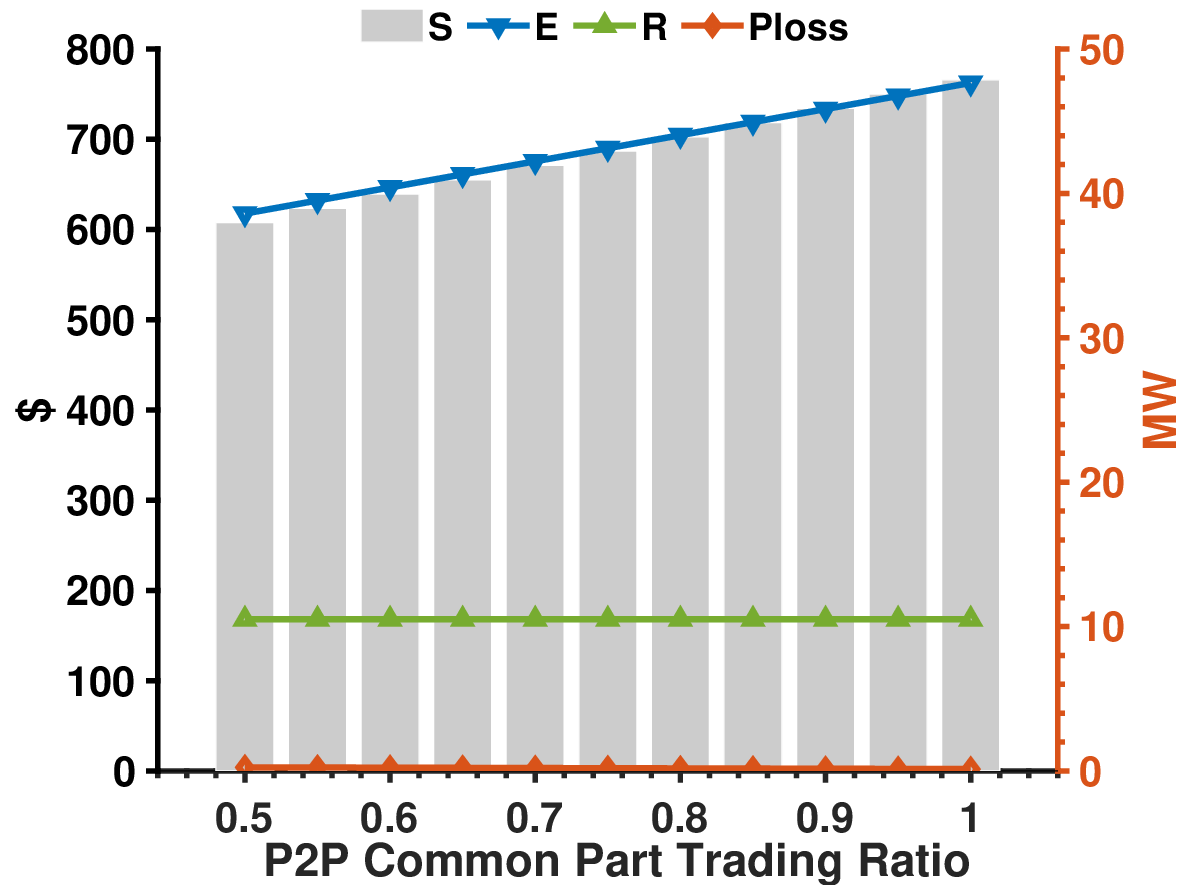}
    \label{(c)}
\end{minipage}}
\centering
  \captionsetup{singlelinecheck=off}
\caption{Simulation results in UKGDS 95-bus system: (a) Under different power contract P2P trading ratios (hour 11), (b)  Under different energy contract P2P trading ratios (hour 15) and (c) Under different P2P common part trading ratios (hour 11)}
\vspace{-0.3cm}
\end{figure*}

\begin{figure*}[htbp]
\centering
\subfigure{
\begin{minipage}[t]{0.38\linewidth}
\centering
\includegraphics[width=\linewidth]{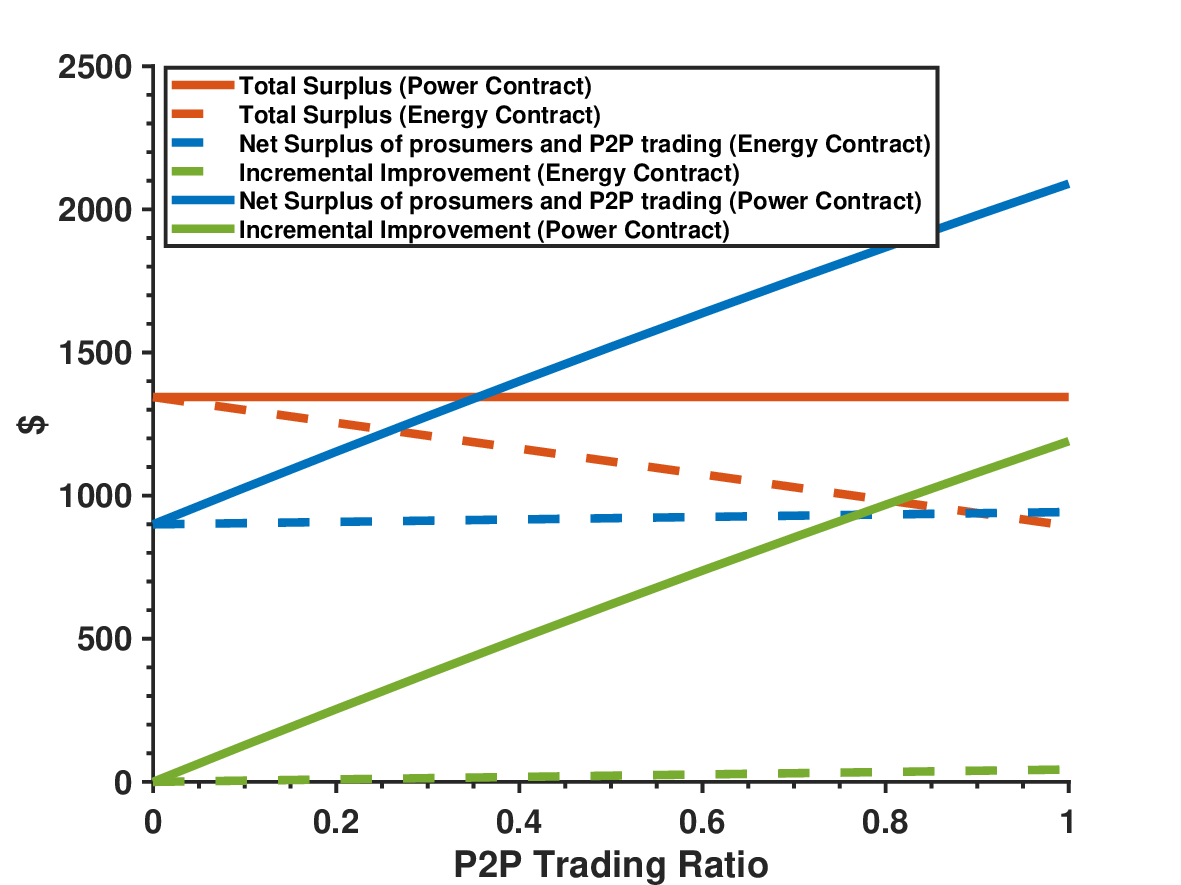} 
    \label{(a)}
\end{minipage}}
\subfigure{\begin{minipage}[t]{0.38\linewidth}
\centering
    \includegraphics[width=\linewidth]{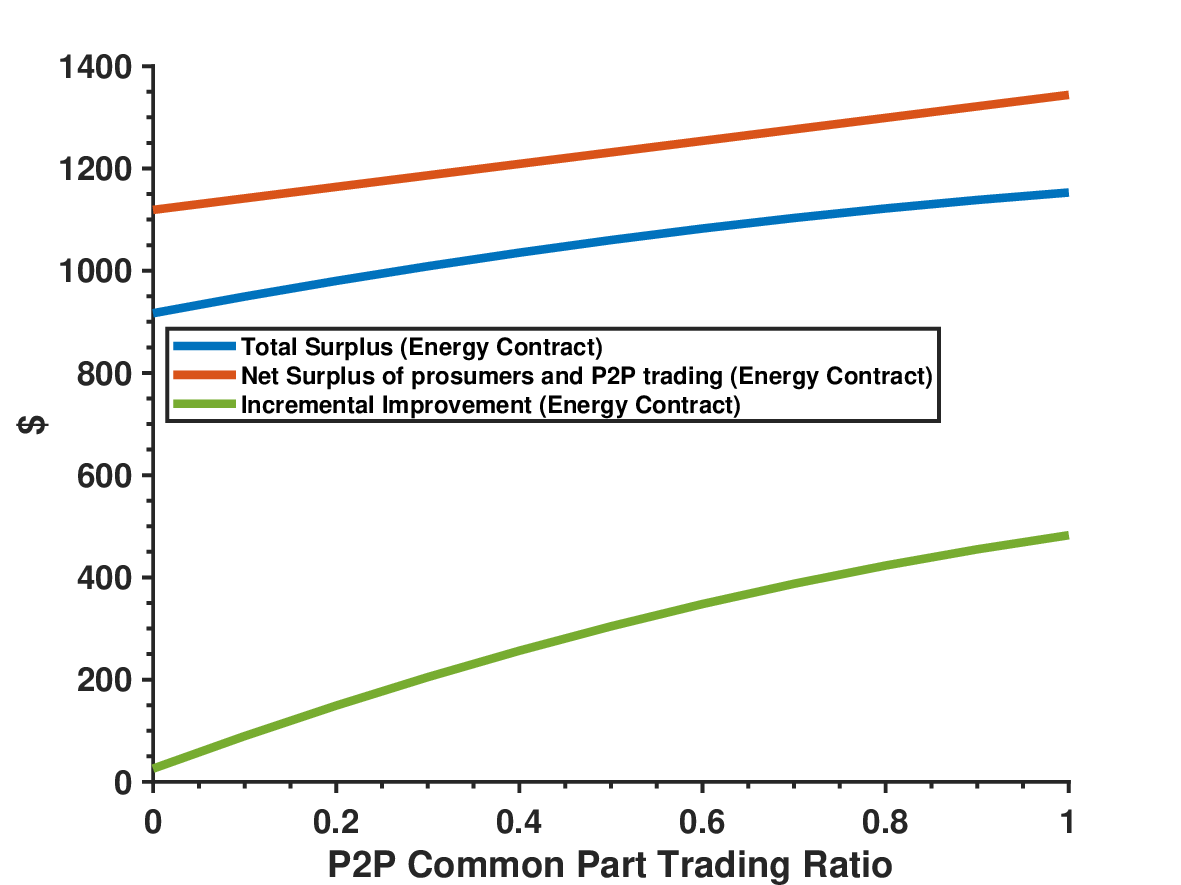}
        \label{(b)}
\end{minipage}}
\centering
  \captionsetup{singlelinecheck=off}
\caption{Economic indices under different (a) P2P trading ratios and (b) P2P common part trading ratios}
\vspace{-0.3cm}
\end{figure*}

\begin{figure*}[htbp]
\centering
\subfigure{
\begin{minipage}[t]{0.36\linewidth}
\centering
\includegraphics[width=\linewidth]{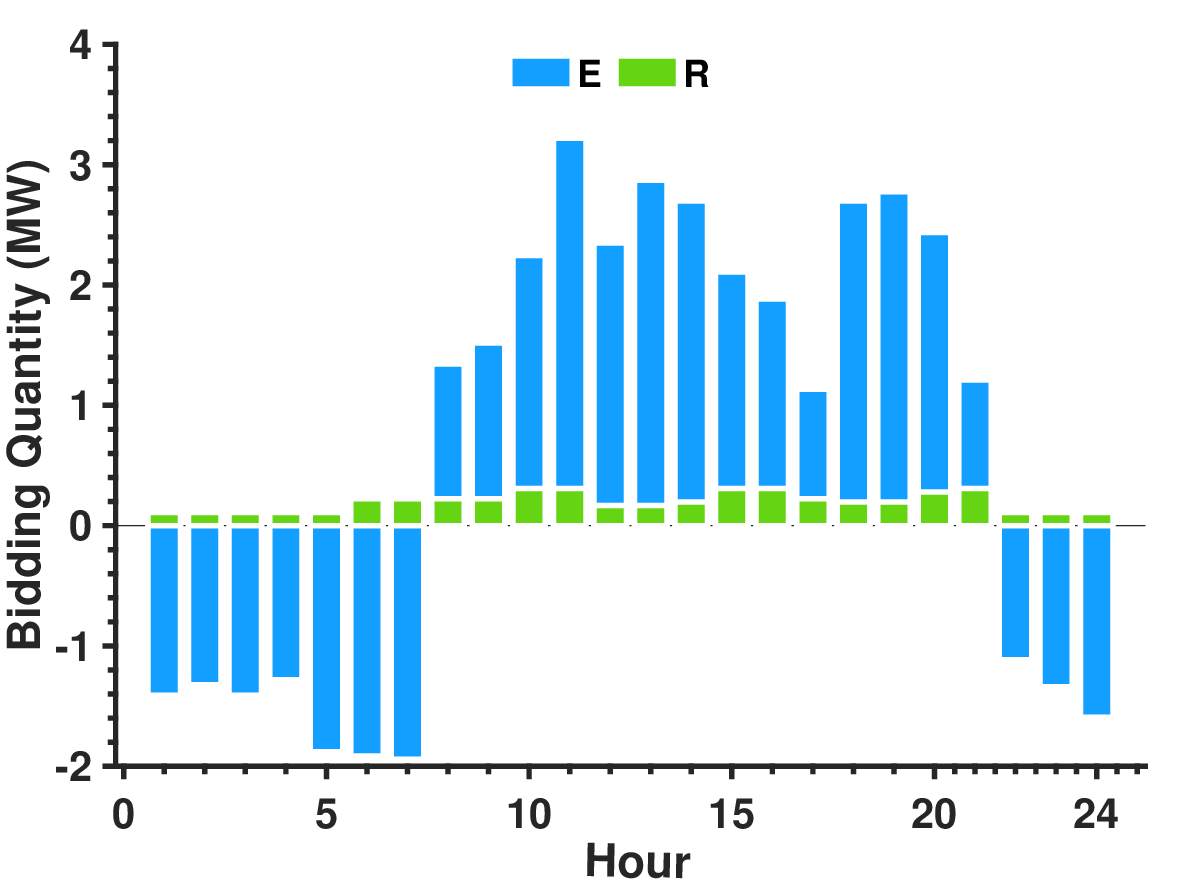} 
    \label{(a)}
\end{minipage}}
\subfigure{\begin{minipage}[t]{0.36\linewidth}
\centering
    \includegraphics[width=\linewidth]{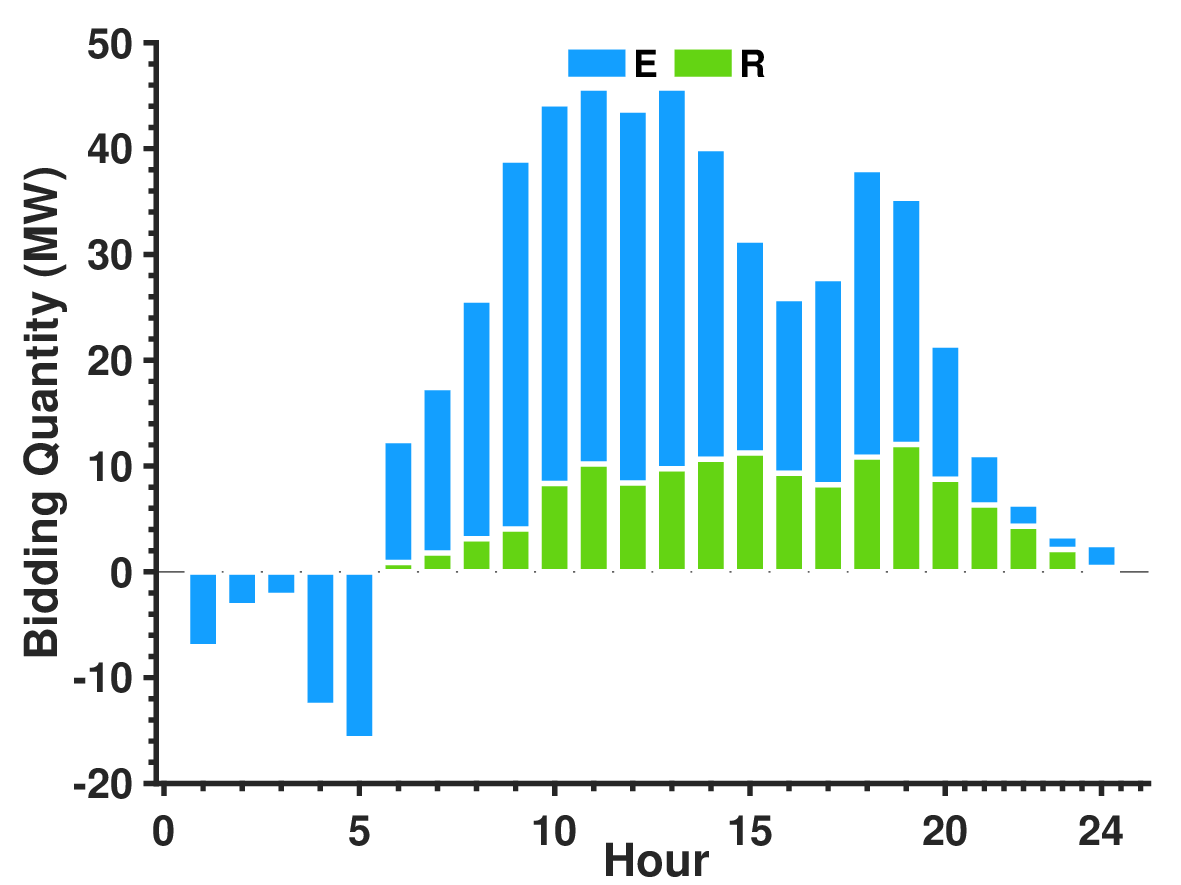}
        \label{(b)}
\end{minipage}}
\centering
  \captionsetup{singlelinecheck=off}
\caption{Energy and reserve provided by DSO in the wholesale market in (a) 2-bus system and (b) UKGDS 95-bus system}
\vspace{-0.5cm}
\end{figure*}
The UKGDS 95-bus system is fed by a 33/11 kV substation transformer\cite{8419280}. There are eighteen prosumers, whose locations are shown in Fig. 7. DGs at nodes 28, 61 and 83 have a generation output range of (0.3-5) MW; remaining DGs have a generation output range of (0.15-4.5) MW. Different types of PV panels are owned by prosumers 7, 9, 16, 26, 31, 77 and 84. Besides, prosumers have an average predicted demand from 0.5 to 2 MW during a day, and a part of load (up to 0.45 MW) can be curtailed during hours 10-12 and 14-18. The exchanging power limit at the substation node is lifted to 55MW. The detailed system setting can be found at \cite{unn2020}.

As for submitted quantities and total surplus, results of testing DSO's optimal operation model under different power contract P2P trading ratios at hour 11 and energy contract P2P trading ratios at hour 15, as shown in Fig. 8 (a) and (b), are similar with that of the 2-bus system, obeying the Theorem 1.  

Moreover, we further investigated the impact of different common part trading ratios of the energy contract P2P trading, which are defined as the ratio of the common part of power between P2P participants to the larger value of P2P seller's output or buyer's demand. According to Fig. 8 (c), as the ratio increases, the energy provided by DSO and total surplus increases because the P2P sellers' PV output is increased and the real-time P2P trading power imbalance decreases, which means that the net power injection of selling prosumers increases and therefore the extent of the DSO compensating the P2P trading power mismatches is alleviated. The submitted quantity in the reserve market will not change due to the obvious price difference when compared to the energy market. Besides, the system power loss will also increase because the net nodal power injection changes, but the change is too slight to tell from the figure.

Fig. 9 (a) presents the economic indices for changing P2P trading ratios. As discussed in the 2-bus system, power contract P2P trading does not affect the total surplus, but will raise the net surplus of prosumers and P2P trading and incremental improvement. For energy contract P2P trading, the total surplus decreases when P2P trading ratio increases. Fig. 9 (b) shows that when the common part trading ratio increases, the total surplus increases because the submitted quantity in the energy market recovers, as well as the net surplus of prosumers and P2P trading and incremental improvement. The reason why the net surplus of prosumers and P2P trading and incremental improvement growths in a quadratic form is that prosumers' utility of consuming P2P trading quantities is a concave function. Note that when the ratio reaches to 1, the energy contract is the same as the power contract P2P trading.

Besides, the 24 hour energy and reserve provided by DSO in the wholesale market of two test systems are illustrated in Fig. 10. It highlights that the prosumers’ participation in the energy market through DSO, whether as a producer or a consumer, primarily depends on the market price and prosumers' operation costs. DSO tends to purchase from the energy market when the price falls below its operation costs. Conversely, in other hours DSO prefers to provide the energy to the market for higher revenues. As the price of reserve market is way lower than that of the energy market, the amount of upward reserve provided by DSO is limited. Last but not least, relaxation gaps of SOCP are examined to ensure the accuracy of simulation results. The maximum relaxation error between two test systems across all tested scenarios is $5\times 10^{-4}$. \vspace{-0.5cm}

\section{Conclusion}
This paper proposes an optimal operation model for a DSO formed by prosumers considering their P2P trading. The optimal operation model aims to decide the optimal dispatching of prosumers' DERs considering network operation limits, achieve load balance, optimize total surplus through providing energy and reserve in the wholesale market. The paper further analyzes the impact of P2P trading on the DSO's optimal operation. The theorem, validated through theoretical proofs and simulations, demonstrates that the energy contract P2P trading will affect quantities of energy exchanged between the DSO and the wholesale market, but not internal dispatch decisions of the DSO. Different levels of real-time power consistency may also lead to different total surpluses in the distribution network. As a ideal case of energy contract P2P trading, power contract P2P trading has no impact on the exchanged energy and total surplus. Therefore, this paper draws two clear-cut conclusions: i) DSO when deciding the optimal operation strategy for the generation demand balance of all prosumers, can safely disregard P2P trading among prosumers, when the amount of power buying and selling of each P2P transaction at any time is the same; ii) P2P trading will not affect physical power flows of the whole system, but it may affect the exchanged energy between the DSO and the wholesale market, and then the financial distribution between the DSO, representing all prosumers, and the prosumers engaged in P2P trading.

\ifCLASSOPTIONcaptionsoff
  \newpage
\fi



%
    \bibliographystyle{ieeetr}    
 \bibliography{lib_test} 
\end{document}